\documentclass[11pt,a4paper]{article}
\pdfoutput=1
\usepackage{jcappub}

\usepackage{bm}
\usepackage{amsfonts}
\usepackage{placeins}
\usepackage{siunitx}
\usepackage{fixme}
\usepackage{todonotes}

\newcommand{\be}{\begin{equation}}
\newcommand{\ee}{\end{equation}}
\newcommand{\bea}{\begin{eqnarray}}
\newcommand{\eea}{\end{eqnarray}}

\newcommand{\unitvec}[1]{{\hat{\bm #1}}}

\newcommand{\fabs}[1]{\left| #1 \right|}

\newcommand{\di}[1]{\text{#1}}
\renewcommand{\vec}[1]{{\bm #1}}

\newcommand{\specialcell}[2][c]{\begin{tabular}[#1]{@{}c@{}}#2\end{tabular}}

\usepackage{multirow}
\usepackage[normalem]{ulem}

\makeatletter
	\newcommand{\vast}{\bBigg@{8}}
\makeatother

\begin{document}



\title{Measuring the velocity field from type Ia supernovae in an LSST-like sky survey}

\author[a]{Io Odderskov,}
\author[a]{Steen Hannestad}

\affiliation[a]{Department of Physics and Astronomy \\ University of Aarhus, Ny Munkegade, Aarhus C, Denmark}

\emailAdd{isho07@phys.au.dk, sth@phys.au.dk}

\abstract{In a few years, the Large Synoptic Survey Telescope will vastly increase the number of type Ia supernovae observed in the local universe. This will allow for a precise mapping of the velocity field and, since the source of peculiar velocities is variations in the density field, cosmological parameters related to the matter distribution can subsequently be extracted from the velocity power spectrum. 
	
One way to quantify this is through the angular power spectrum of radial peculiar velocities on spheres at different redshifts. We investigate how well this observable can be measured, despite the problems caused by areas with no information. To obtain a realistic distribution of supernovae, we create mock supernova catalogs by using a semi-analytical code for galaxy formation on the merger trees extracted from N-body simulations. We measure the cosmic variance in the velocity power spectrum by repeating the procedure many times for differently located observers, and vary several aspects of the analysis, such as the observer environment, to see how this affects the measurements. 
	
Our results confirm the findings from earlier studies regarding the precision with which the angular velocity power spectrum can be determined in the near future. This level of precision has been found to imply, that the angular velocity power spectrum from type Ia supernovae is competitive in its potential to measure parameters such as $\sigma_8$. This makes the peculiar velocity power spectrum from type Ia supernovae a promising new observable, which deserves further attention.}
\maketitle

\section{Introduction}

In the near future, the Large Synoptic Survey Telescope (LSST) will carry out observations of millions of type Ia supernovae \cite{LSST2009}. As the best standard candles at cosmological distances, the lightcurves of these supernovae will provide precise luminosity distances to millions of galaxies throughout the universe. For some sub-sample of these, the spectroscopic redshift will be measured for the host galaxy, and together, the supernova luminosity distances and spectroscopic redshifts will lead to an unprecedented opportunity for describing the velocity structure of the universe. Such observations can for example be used to measure the expansion history, and provide precise measurements of the Hubble constant, greatly improving the already extremely successful measurements that have been done in the past \cite{Perlmutter1999,Riess1998,Schmidt1998,Riess2016}.

However, the observed redshifts and luminosities do not only carry information about the expansion history. For one thing, the light we receive is lensed and redshifted by the gravitational fields of structures it has passed through \cite{Hui2006,Bonvin2006}. Such effects are negligible for light sources at small distances, as the light we receive from these has not passed through as many structures. But in this regime the peculiar velocities -- typically a few $\SI{100}{km/s}$ -- account for a significant fraction of the measured redshift. These effects mean that neither the redshifts nor the luminosity distances measured from type Ia supernovae are equal to the values they would have in a homogeneous and isotropic universe. And since both peculiar velocities and the gravitational effects on light are induced by large scale structure, the imprints they make on the supernovae are correlated, and therefore needs to be taken into account to avoid making biased parameter estimates \cite{Hui2006,Sugiura1999,Neill2007,Davis2010}.

But the imprint from large scale structure on luminosity distances and redshifts can also be used as a signal in itself. In \cite{Bonvin2006,Bonvin2006b}, it was explored how the correlations in the luminosity distances of distant supernovae can be used to measure cosmological parameters. 
In this paper, we will instead consider the potential for using observations of type Ia supernovae at low redshift as a cosmological probe. By assuming a specific expansion rate, the luminosity distances and redshifts can be used to determine the peculiar velocities of the observed galaxies \cite{Davis2014}. And since these are created by the gravitational attraction of over-densities, they can be used to probe the matter distribution in the local universe and its associated parameters \cite{Gordon2007,Hannestad2007,Watkins2007,Irsic2011}.

One way to describe the peculiar velocity field is through a multipole expansion of the radial peculiar velocities. With the current sample of type Ia supernovae, only the lowest multipoles can be reliably estimated \cite{Weyant2011,Haugboelle2007}. But this will change in the near future with the large sample of type Ia supernovae collected in the LSST sky survey.

In this paper, we investigate the potential for a sky survey with the geometry of the LSST main survey for measuring the angular power spectrum of the radial peculiar velocity field from type Ia supernovae. The basic idea is the same as presented in \cite{Hannestad2007}, but we use a more realistic distribution of supernovae for our mock surveys, as well as the actual geometry of the LSST sky survey. This allows us to study the problems that arise due to incomplete sky coverage and the non-uniform distribution of supernovae in the sky. We use methods developed for analysis of the CMB to correct for the mixing of different multipoles with each other, which is caused by large, unobserved areas in the survey geometry. And we smooth the velocity field to remove features on scales comparable to the size of the largest holes in the distribution of supernovae within the survey. In order to get a realistic prediction for where type Ia supernovae occur, we run a large N-body simulation and populate it with type Ia supernovae based on the star formation history in the individual halos which we obtain with a code for semi-analytical galaxy formation. 

The structure of the paper is as follows: In section \ref{sec:VelocityField}, we give a short description of the connection between the matter distribution and the peculiar velocity field from linear perturbation theory. The simulations and methods for constructing the supernova catalogs are described in section \ref{sec:MockCatalogs}, and the selection of observers and mock observations are described in section \ref{sec:ObserversAndObservations}. In section \ref{sec:AngularPowerSpectrum}, we describe how the angular power spectrum is measured, and how it is affected by the smoothing process and by the correction for the coupling of multipoles caused by the missing sky coverage. We present our results in section \ref{sec:Results} and conclude in section \ref{sec:Conclusions}.

\section{The power spectrum of the peculiar velocity field}
\label{sec:VelocityField}

The theoretical predictions for the peculiar velocity field are provided by gravitational instability theory, which describes how structures grow in response to small perturbations in the density field (see for example \cite[chapter 14]{Peebles1980}). The comoving velocity field, $\vec{v}(\vec{x})=\dot{\vec{x}}$, describes the peculiar velocity of a mass element at each position, and the density field is customarily described using the over-density parameter, $\delta(\vec{x})=\frac{\rho(\vec{x})-\bar{\rho}}{\bar{\rho}}$, where $\bar{\rho}$ is the mean density of the universe, and $\rho(\vec{x})$ is the density at position $\vec{x}$. According to gravitational instability theory, these fields are related via the continuity equation, $\frac{\partial \delta}{\partial t}+\nabla \cdot \vec{v} = 0$.\footnote{Note that the perturbations in the fields need to be small enough to be described by linear perturbation theory, i.e. both the density and the velocity field need to be smoothed over non-linear scales.} 

To obtain an expression for how the peculiar velocity field depends on cosmological parameters, we first express the time derivative of the density in terms of the growth function, $f=\frac{d\log \delta}{d\log a}$, as $\frac{\partial \delta}{\partial t} =  \frac{\dot{a}}{a} f \delta = H f \delta$. Using this and Fourier transforming the continuity equation, it is found that the Fourier component of the velocity field stemming from density perturbations of wavelength $\frac{2\pi}{k}$ and direction $\hat{\vec{k}}$ is  
\begin{align}
\vec{v}_{\vec{k}}(a) = -\frac{ifH\delta_{\vec{k}}(a)}{k} \hat{\vec{k}}. 
\label{eq:vk}
\end{align}
The occurrence of $k$ in the denominator shows that the velocity field is more sensitive to clustering on large scales than studies of the density field. This means that  analyses of the velocity field makes it possible to study clustering on larger scales than can be reached using direct measurements of the matter distribution.

From equation \ref{eq:vk}, the power spectrum of the perturbations of the velocity field, $P_v(k)\equiv \langle |\vec{v}_{\vec{k}}|^2 \rangle$, is found to be related to the matter power spectrum, $P_m(k)\equiv \langle |\delta_{\vec{k}}|^2 \rangle$, via:
\begin{align}
P_v(k) = H^2 f(\Omega)^2 k^{-2} P_m(k).
\end{align}
The close relationship between $P_m(k)$ and $P_v(k)$ implies that parameters associated with the matter power spectrum, such as its shape and amplitude, can be determined from studies of the velocity field. If all three components of the peculiar velocities could be observed, the equation above relating the 3D power spectra of density fluctuations and of the velocity field could be directly used for parameter estimation. However, this is not the case; since the velocities are measured from the redshift of the observed galaxies, only the radial component can be measured. For this reason, a more suited observable than the 3D power spectrum is the 2D power spectra of the radial peculiar velocities in shells at different redshifts, which is therefore the observable considered in this study. 

\section{Construction of mock supernova catalogs}
\label{sec:MockCatalogs}

The potential for measuring the peculiar velocity field from observations of type Ia supernovae is sensitive to the distribution of these events, since the peculiar velocity can only be measured at positions where there actually is a supernova to observe. Therefore, studies such as this one must be based on accurate mock catalogs, realistically accounting for where type Ia supernovae occur in the cosmic web. In this section, we describe how the type Ia supernova catalogs are constructed. 

\subsection{Simulations}

The large scale structure of the universe is best captured by N-body simulations, which therefore form the basis for our analysis. The simulations are based on cosmological parameters in agreement with those determined from the CMB by the Planck collaboration \cite{Planck2013}. We use a modified version of the GADGET-2 code \cite{Springel2001,Springel2005},
with initial conditions generated using a code written by J. Brandbyge
\cite{Brandbyge2010} based on transfer functions computed using
CAMB\footnote{\url{http://camb.info/}} \cite{Lewis2002}. Specifically, the transfer functions are calculated
with $(\Omega_{b},\Omega_{CDM})=(0.048,0.26)$, whereas only cold
dark matter is used in the N-body simulations. A flat universe is
assumed, and $(h,\sigma_{8})=(0.68,0.85)$. The simulations are run
from a redshift of $z=50$ until $z=0$.

The main simulation is run in a periodic box of side length $\SI{512}{Mpc/h}$. In order to determine the necessary resolution to resolve the galaxies in which type Ia supernovae occur, we also run a set of smaller simulations in boxes of side length $\SI{64}{Mpc/h}$ with varying mass resolutions, as summarized in table \ref{tab:simulations}.
\begin{center}
	\begin{table}
		\center  
		\begin{tabular}{lccccc}
			\hline\hline
			\rule{0pt}{3ex} &$N_\textrm{part}$ & Box size $[\si{Mpc/h}]$ & $M_\textrm{min}\, [\num{e10}\di{M}_\odot/\di{h}]$ &$N_\textrm{snaps}$ \\
			\hline 
			\rule{0pt}{3ex}
			\multirow{7}{*}{Test \vast\{ }
			& $128^3$   & $64$ &$21$ & $45$ \\
			&$174^3$  & $64$ &$8$ & $45$ \\
			\rule{0pt}{4ex} 
			&$200^3$        &      $64$ &$6$ & $30$  \\
			&$200^3$  &$64$ &$6$  &$45$  \\
			&$200^3$            & $64$ &$6$ & $60$  \\
			\rule{0pt}{4ex}
			&$224^3$  &$64$ &$4$  &  $45$ \\
			&$312^3$  & $64$ &$1$ & $45$ \\
			\rule{0pt}{4ex}
			Main &$1600^3$  &$512$ & $6$ & $45$ \\
			\hline
		\end{tabular}
		\caption{Table over the spatial and temporal resolutions of the simulations. According to the ROCKSTAR user manual \cite{RockstarManual}, the halo finder can be trusted down to halos comprised of 20 DM particles. We show which mass this corresponds to in each of the simulations.} 
		\label{tab:simulations}
	\end{table}
\end{center}

As will be discussed in section \ref{sec:SAM}, we find that a sufficient resolution for our main simulation is obtained with $1600^3$ particles in the box. The temporal resolution is found not to make any significant difference for the galaxy catalogs. For the main simulation, we make snapshots at 45 different times, linearly separated in scale factor from the beginning of the simulation to now. 

\subsection{Construction of merger trees}

To generate halo catalogs for the simulations described above, we use the halo finder ROCKSTAR \cite{Behroozi2011}, and subsequently obtain the merger history of the halos by using the Consistent Trees code \cite{Behroozi2012}. 

ROCKSTAR identifies halos by using a variant of the Friends-of-Friends (FOF) algorithm. At first, particles that are close together in space are grouped together using the FOF algorithm, which assembles particles that are within
a specified distance of each other. Within each FOF-group, the phase-space distances between individual particles are computed from a metric that combines differences in position and velocity into a single measure. Using this, locations in which the mean phase-space-distance between particles is low are identified as local maxima of phase-space density.
These maxima are used as seed halos and all particles in the original FOF-group are assigned to the seed that is closest in phase-space. At last, halos located in more massive hosts are categorized as subhalos, unbound particles are removed, and halo masses are calculated as spherical overdensities. For each halo, the position of the halo centre is calculated
as the mean of the positions of the central particles, and the halo velocity is calculated as the mean of the velocities of the innermost 10\% of the halo particles. 

When the halo catalogs have been generated for all the snapshots for a given simulation, these are fed to the Consistent Trees code. This code constructs merger trees by making predictions of the halo properties, such as position and velocity, backwards in time, and thereby identifies the most likely progenitor of a given halo. Halos for which no consistent history can be identified are removed from the catalogs. In this way, reliable and consistent merger trees are obtained, which can then be used to predict the formation and evolution of galaxies as described in the next section.

\subsection{Semi-analytical identification of galaxies}
\label{sec:SAM}

The distribution of supernovae throughout the universe is dictated by the mechanisms of star formation. Even though this is not a completely understood field, it can be imitated
using semi analytic galaxy formation models (SAMs), consisting of several different components for modelling different aspects of the star- and galaxy formation. We use the publicly available semi-analytical code Galacticus \cite{Benson2012}, which we use in the so-called "hybrid" mode, in which the dark matter halo merger trees are extracted from an
N-body simulation, and only the baryonic parts of galaxy and star formation are handled in a semi-analytical manner.

The halo merger tree form the basic structure which determines when and where galaxies form. After recombination, 
gas is assumed to fall into the gravitational potential wells of the dark matter halos. In this process, the gas is heated, and as it subsequently cools, it falls to the centers of the halos where it forms disk galaxies. Elliptical galaxies are predominantly formed when large galaxies merge with each other \cite{Baugh2006}.

The galaxy formation processes are followed by Galacticus by treating the halos of the merger tree as a set of linked nodes containing a number of different components, such as a stellar disk, spheroid, and hot gas halo, which each have various properties. Galacticus evolves the properties of these components forward in time subject to a collection of differential equations, which describe their evolution, and rules for what happens in the case of various events, such as two galaxies merging with each other.

In addition to outputting the positions, velocities and other properties of the galaxies at different times, the star formation history of each node can be recorded, as will be described in the next section.

\begin{figure}[h!]
	\centering
	\includegraphics[width=0.49\textwidth]{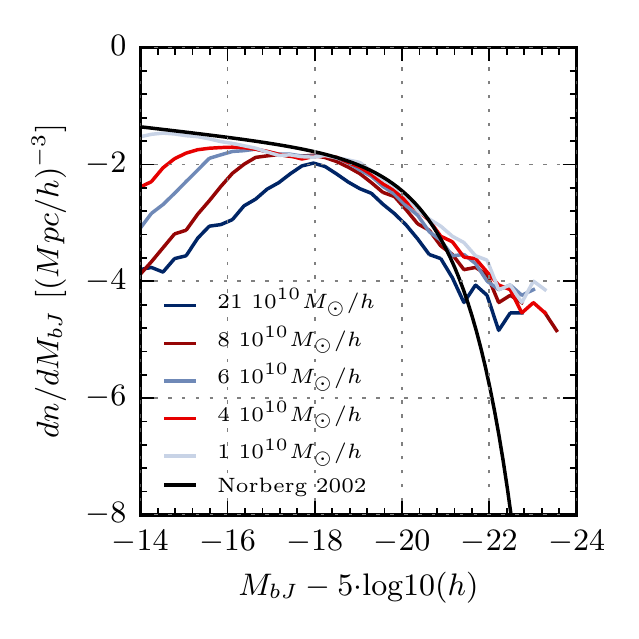}
	\includegraphics[width=0.49\textwidth]{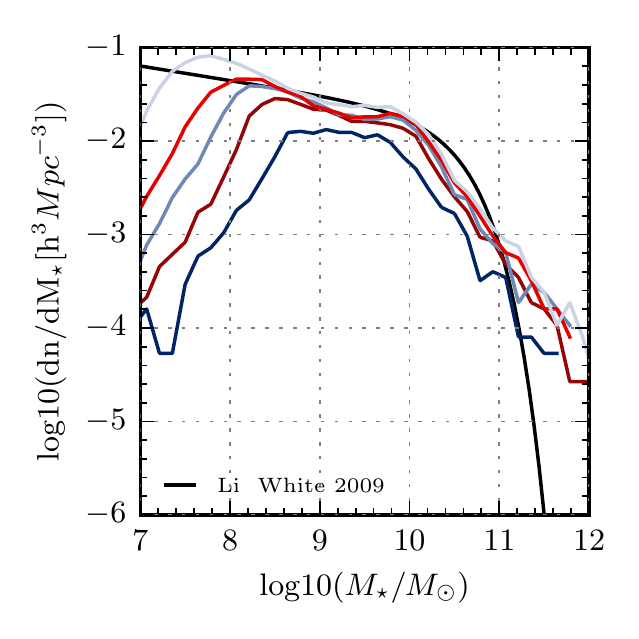}
	\includegraphics[width=0.49\textwidth]{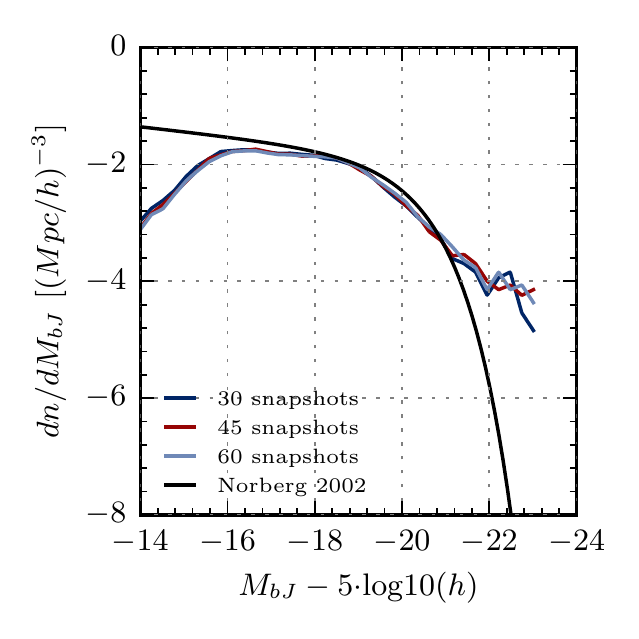}
	\includegraphics[width=0.49\textwidth]{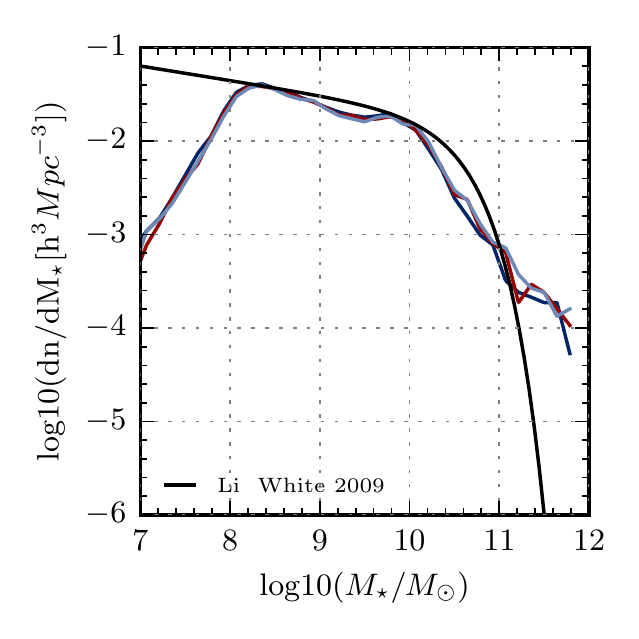}
	\caption{In order to determine the necessary resolution in mass and time for the N-body simulation, we compare the luminosity function and stellar mass function in the test simulations to observations. Specifically, we compare to the luminosity function of 2dF galaxies \cite{Norberg2002}, and to the stellar mass function of SDSS galaxies \cite{Li2009}. The luminosity functions are shown on the \textbf{left panel}, and the stellar mass functions on the \textbf{right panel}. In the \textbf{top panel}, the simulations of different mass resolutions are compared. The resolutions are shown in the legend of the figure to the left, as the approximate mass of the least massive halo that can be identified in the given simulation (consisting of 20 particles).  It is evident that increasing the resolution of the simulation affects the formation of small galaxies. In the \textbf{bottom panel}, the simulations of different temporal resolution are compared. This is found to make a negligible difference.}
	\label{fig:LFs_and_stellarMassFunctions}
\end{figure}

\begin{figure}[h]
	\centering
	\includegraphics[width=0.49\textwidth]{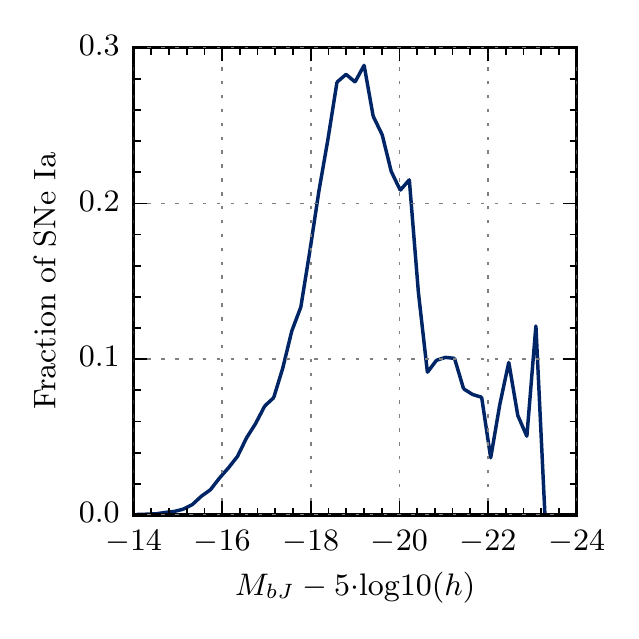}
	\includegraphics[width=0.49\textwidth]{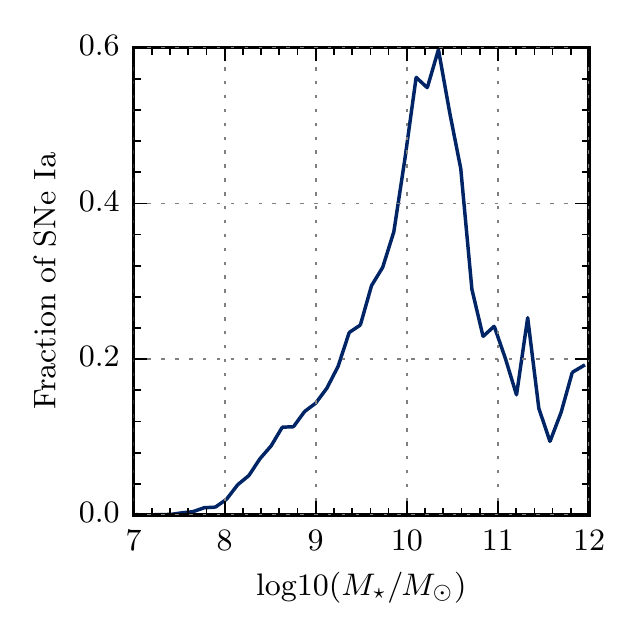}
	\caption{The frequency of supernovae as a function of galactic magnitude (\textbf{left}) and stellar mass (\textbf{right}), in a simulation of box size $\SI{64}{Mpc/h}$ and $200^3$ particles. It is seen that most of the supernovae occur in galaxies with $M_{b_J}-5\log h > -21$, and a stellar mass less than $\num{e11}\,\di{M}_\odot$. The total type Ia supernova rate in the box is $\SI{1.8e-4}{yr^{-1} h^3 Mpc^{-3}}$.
		}
	\label{fig:rate_vs_StellarMass_and_Magnitude}
\end{figure}

In the top of figure \ref{fig:LFs_and_stellarMassFunctions}, the luminosity functions and stellar mass functions of the galaxies identified by Galacticus in the five small test simulations of different spatial resolutions are shown. We examine in which galaxies most of the supernovae occur by calculating the present day type Ia supernova rate in each galaxy using the method described in the next section. This is shown for one of the simulations in figure  \ref{fig:rate_vs_StellarMass_and_Magnitude}, which shows histograms over the absolute magnitudes and stellar masses of the galaxies, each weighted by its present day type Ia supernova rate. We find that type Ia supernovae tend to occur in galaxies with absolute magnitude in the $b_J$-band obeying $M_{b_J}-5\log h < -16$ and a stellar mass greater than $\num{e8}\,\di{M}_\odot$. As can be seen in figure \ref{fig:LFs_and_stellarMassFunctions}, such galaxies are well resolved in the simulation which has a minimal halo mass of $\num{6e10}\,\di{M}_\odot/\di{h}$, corresponding to a mass resolution of $1600^3$ particles in the box of side length $\SI{512}{Mpc/h}$. In the bottom of figure \ref{fig:LFs_and_stellarMassFunctions}, the comparison of simulations with different temporal resolutions is shown, revealing that this does not have any significant effect. We choose a temporal resolution of 45 snapshots, linearly spaced in scale factor from the beginning of the simulation at $z=50$ to $z=0$ for the simulation for the main analysis.

There appears to be too many galaxies with a high luminosity and a large stellar mass. The luminosity function has not been corrected for extinction of light by gas in the galaxies, which could partly explain the discrepancy between the measured and the predicted number of very luminous galaxies. As can be seen from figure \ref{fig:rate_vs_StellarMass_and_Magnitude}, we find that the majority of type Ia supernovae comes from galaxies with magnitudes of $M_{b_J}-5\log h > -21$ and stellar masses below $\num{e11}\,\di{M}_\odot$. Therefore, we do not expect the over-representation of massive, luminous galaxies to have a significant effect on our results.

\subsection{Star formation histories and type Ia supernovae}
\label{sec:SFH}

\begin{figure}[h]
	\centering
	\includegraphics[width=\textwidth]{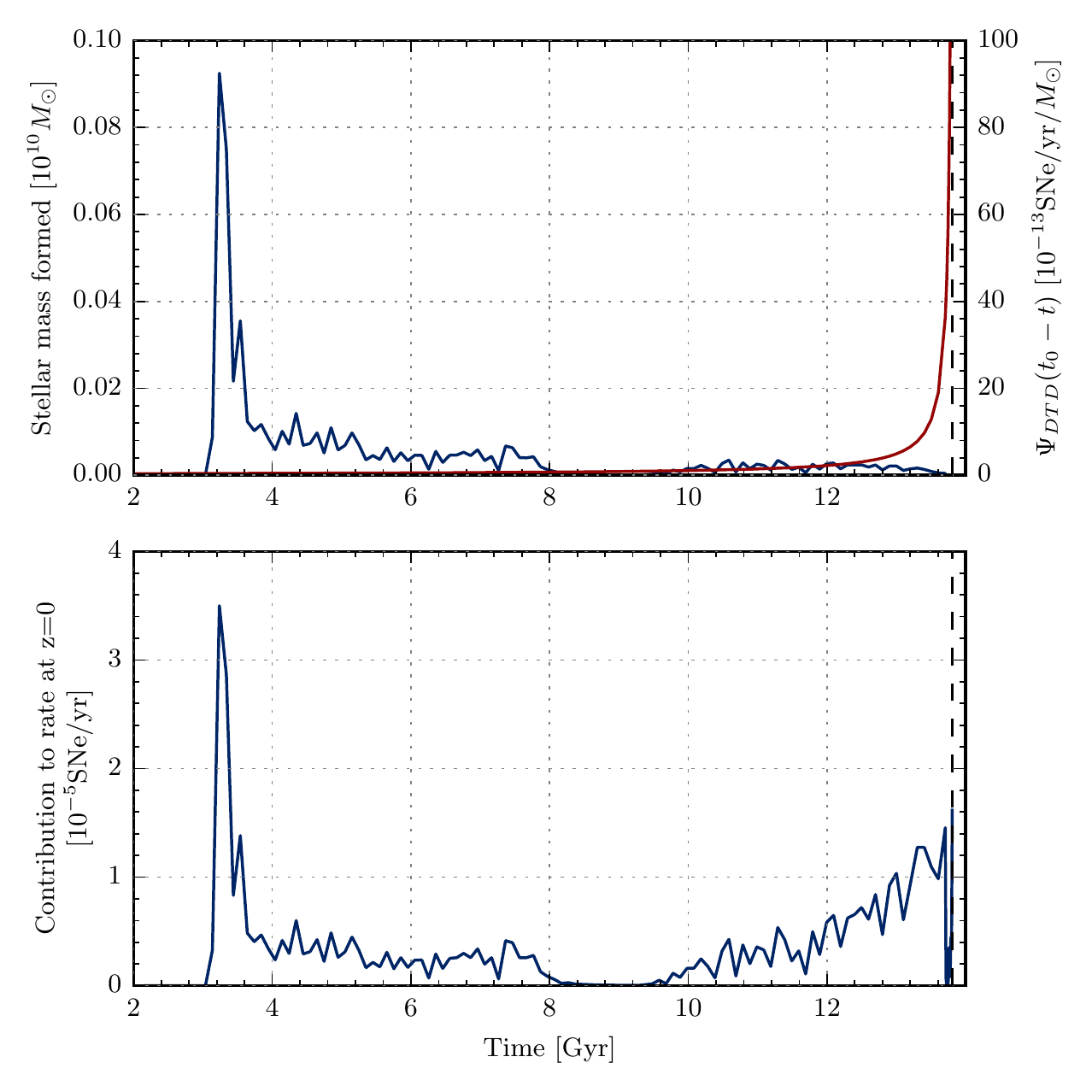}
	\caption{\textbf{Top:} Example of a star formation history; the stellar mass formed at each time is shown on the left axis (full blue line) as a function of time $t$. This is plotted together with the delay time distribution (red line and right axis). The black, dashed line marks $t=t_0$, the present age of the universe. \textbf{Bottom:} Contribution to the present day type Ia supernova rate from the stars formed at each time, obtained by multiplying the star formation history and delay time distribution shown above. By integrating this function, the total present day supernova rate for the galaxy is found, as described in equation \ref{eq:RIa}.}
	\label{fig:SFH_DTD_and_SNRate}
\end{figure}

To predict the present day rate of type Ia supernovae in each of the galaxies identified by Galacticus, we use the delay time distribution formalism, which is described in e.g. \cite{Maoz2011}. In this formalism, the expected rate of type Ia supernovae in a given galaxy is determined from its star formation history. This is done by multiplying the amount of stellar mass formed at each time with the delay time distribution, $\Psi_\textrm{DTD}(\tau)$, which predicts the number of type Ia supernovae that will result from a unit mass of stars which was formed a time $\tau$ ago. The rate of type Ia supernovae in a given galaxy at time $t$, $r_\textrm{Ia}(t)$, can be obtained by convolving the star formation history, $\Psi$, by the DTD, i.e, 

\begin{align}
r_\textrm{Ia}(t) = \int_0^t \Psi(t-\tau)\Psi_\textrm{DTD}(\tau)d\tau.
\label{eq:RIa}
\end{align} 
This is illustrated in figure \ref{fig:SFH_DTD_and_SNRate}. In the top panel, an example star formation history is shown, along with the delay time distribution. In the bottom panel of the figure, it is shown how many supernovae per year is predicted to result at $z=0$ from that specific star formation history, equivalent to the integrand of equation \ref{eq:RIa}. The total rate of type Ia supernovae at $z=0$ is obtained by integrating this function.

In \cite{Maoz2011}, it is found that good agreement with observations is obtained with
\begin{align}
\Psi(t)_\textrm{DTD} = \SI{4e-13}{SN.yr^{-1}}\,\di{M}_\odot^{-1}\left(\frac{t}{\SI{1}{Gyr}}\right)^{-1}.
\end{align}

\noindent
Using this delay time distribution and the star formation histories calculated by Galacticus, we find that the type Ia supernova rates are proportional to the luminosities of the galaxies. This is in agreement with the results presented in \cite{Yasuda2009}, where the luminosity function of the host galaxies of type Ia supernovae in the SDSS survey is examined, and it is found that the type Ia supernovae do not tend to occur in galaxies of a specific type or color, but with a frequency mainly determined by the luminosity of the galaxies. 

The total, volumetric supernova rate obtained for the main simulation that will be used for our analysis turns out to be $\SI{1.1e-4}{SNe.yr^{-1}.h^{3}.Mpc^{-3}}$. This is roughly a factor of $4$ higher than the rate obtained in \cite{Dilday2008}, which is the one quoted in the LSST Science Book \cite{LSST2009}, 
but in very good agreement with the (somewhat more uncertain) rates obtained in \cite{Neill2007SNeIa,Sullivan2006}.

\section{Mock observers and observations}
\label{sec:ObserversAndObservations}

\begin{figure}[h]
	\centering
	\includegraphics[width=0.49\textwidth]{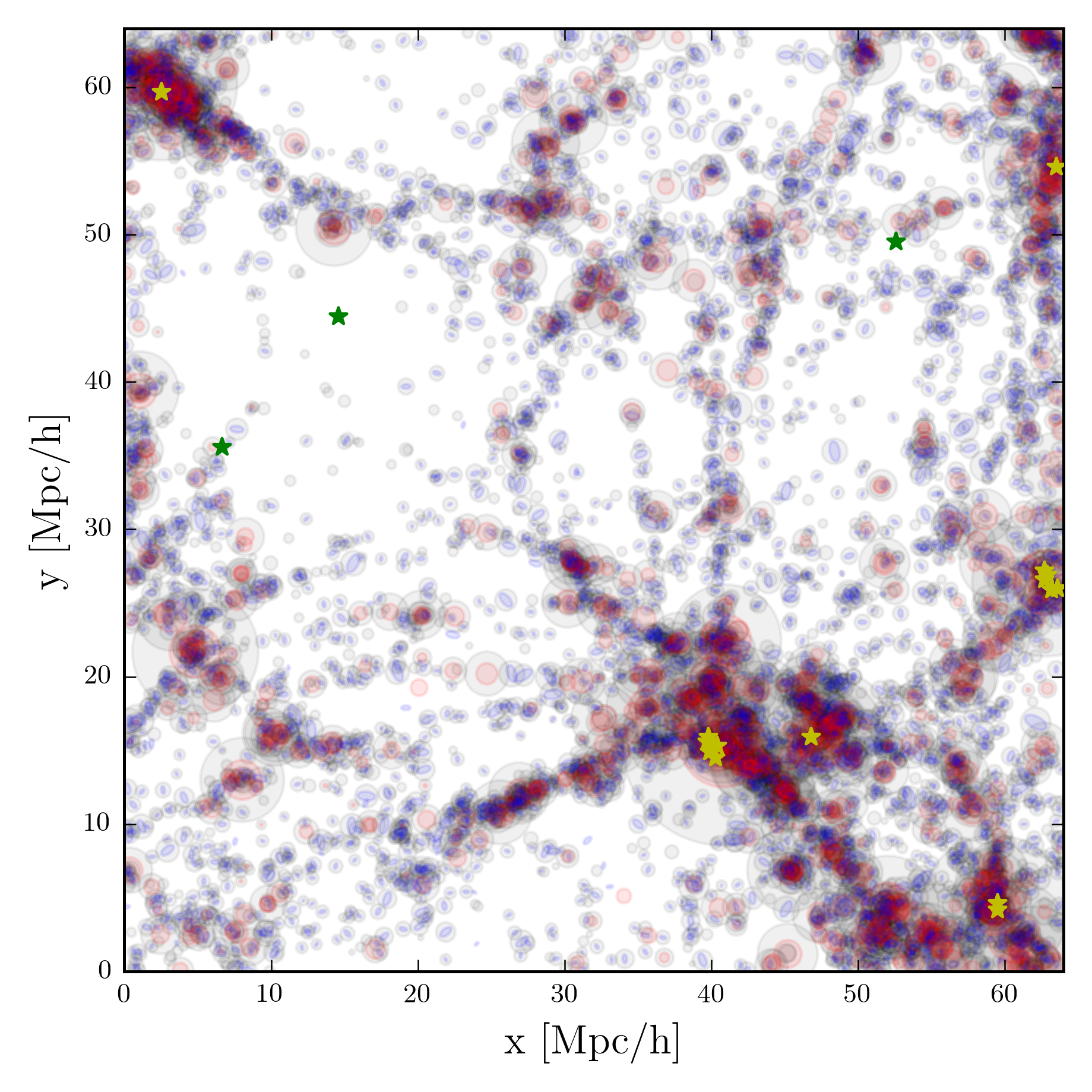}
	\includegraphics[width=0.49\textwidth]{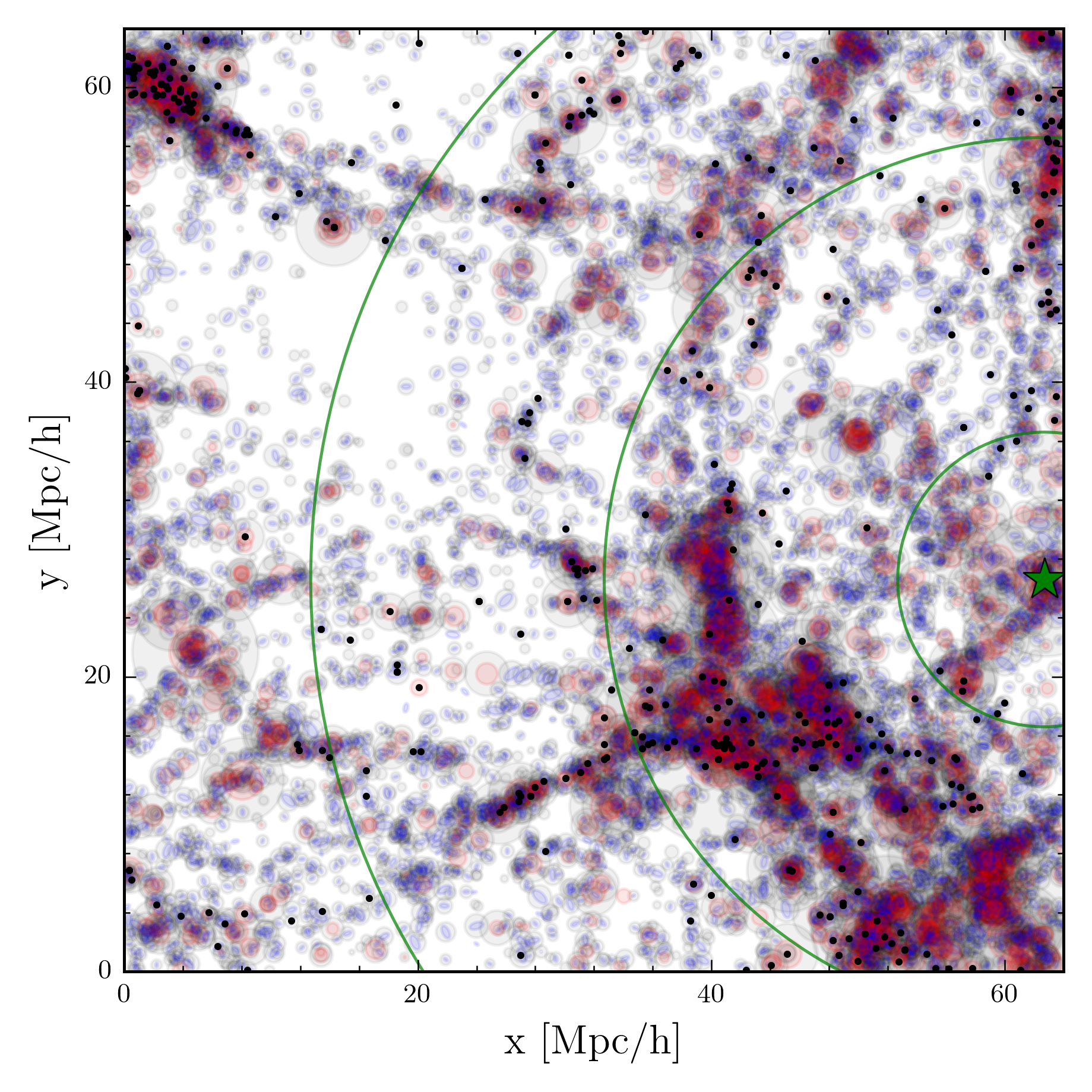}
	\caption{\textbf{Left:} An illustration of the dark matter halos (grey circles), the elliptical galaxies (red circles) and the disk galaxies (blue disks), along with two different sets of observers: The yellow stars mark observers in disk galaxies, located in halos of a mass similar to that of the Virgo Super Cluster, and the green star marks the same number of observers, distributed randomly in the box. The sizes of the halos and galaxies are proportional to the cubic root of their masses, with the disks plotted 5 times greater than the elliptical galaxies and 10 times greater than the dark matter halos. \textbf{Right:} An illustration of the observations (black dots) performed by the observer located at the green star, distributed in bins separated by the green circles.}
	\label{fig:observers_and_observations}
\end{figure}

To investigate how accurately the velocity field can be measured from observations of type Ia supernovae, taking into account cosmic variance, we place a large number of observers at different positions throughout the simulation volume. Each observer identifies sources -- type Ia supernovae and their host galaxies -- out to a maximal distance of $\SI{256}{Mpc/h}$, or $z=0.087$, corresponding to half the sidelength of the simulation box; this is possible for each observer thanks to the periodicity of the simulation. The distributions of observers and observed sources are illustrated in figure \ref{fig:observers_and_observations}. The observers are then assumed to determine the luminosity distances to the observed sources, which can be obtained from the lightcurves of the type Ia supernovae, as well as the spectroscopic redshifts of their host galaxies. The luminosity distances and the measured redshifts can be converted to peculiar velocities by assuming some fiducial expansion history, $H(z)$, as outlined in e.g. \cite{Davis2014}. Specifically, the peculiar velocity of a source with luminosity distance $D_L$ and measured redshift $z$ can be obtained by first calculating its cosmological redshift, i.e. the redshift that would have been measured if the universe had been entirely homogeneous and isotropic universe, $\bar{z}$. This can be determined from the luminosity distance as
\begin{align}
D_L(\bar{z}) = (1+\bar{z})c\int_0^{\bar{z}} \frac{dz}{H(z)}.
\end{align} 
The peculiar velocity can then be calculated from the difference between the cosmological and the actual redshift of the source as
\begin{align}
v_p = c\frac{z-\bar{z}}{1+\bar{z}},
\end{align} 
assuming that the peculiar velocity is non-relativistic.

\begin{figure}[h]
	\centering
	\includegraphics{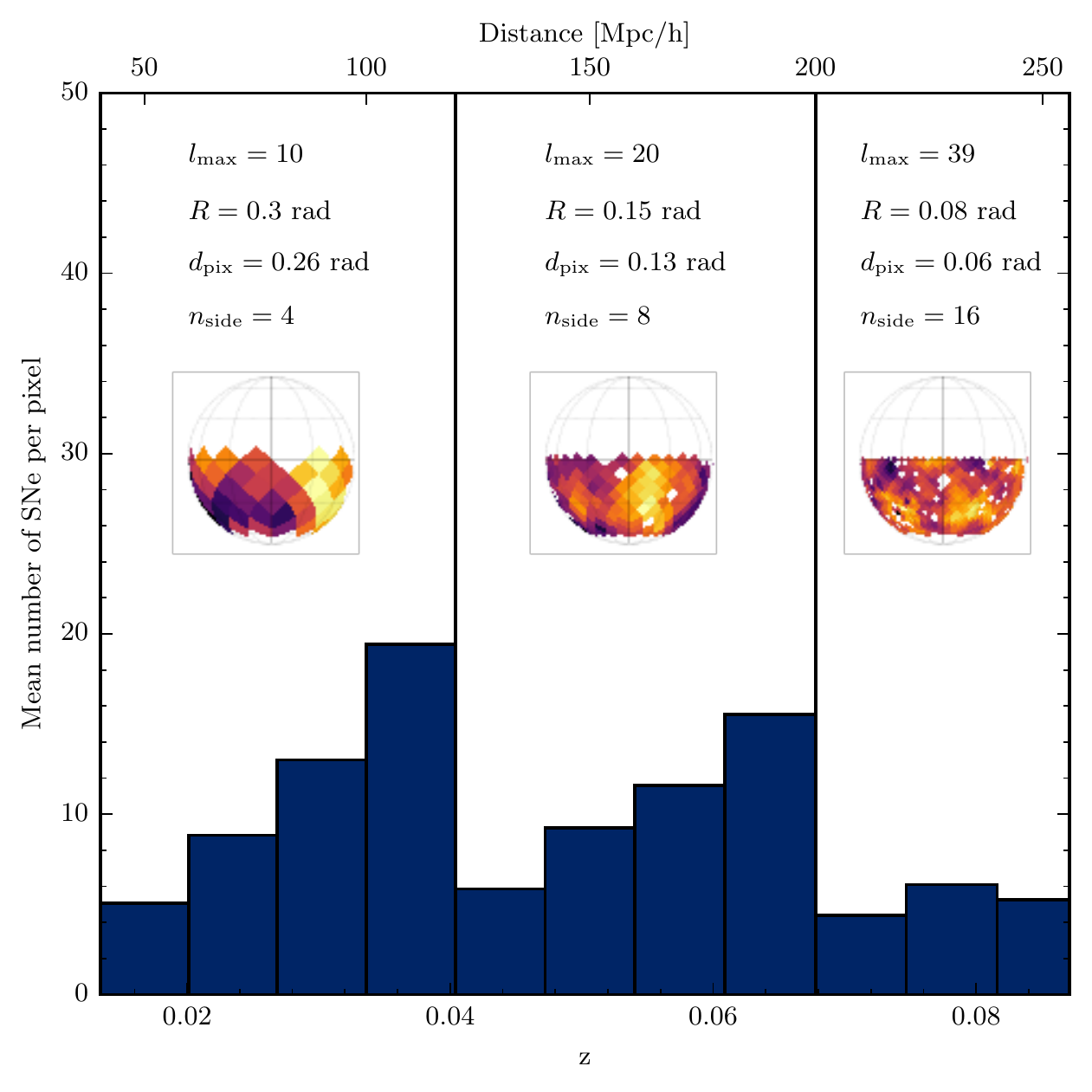}
	\caption{The average number of type Ia supernovae per pixel in each redshift shell. The survey is separated into three regimes, where different resolutions can be obtained. The insets shows an example of a velocity field in a bin from each regime, as an illustration of the pixelation. The velocity field is described by its values in $12\cdot n_\textrm{side}^2$ pixels, with each pixel having a size of $d_\textrm{pix}$. It is smoothed with a Gaussian of FWHM given by $R$, which determines the highest multipole moments that can be measured in each regime, $l_\textrm{max}$.}
	\label{fig:distancebins}
\end{figure}

\begin{figure}[h]
	\centering
	\includegraphics[width=0.7\textwidth]{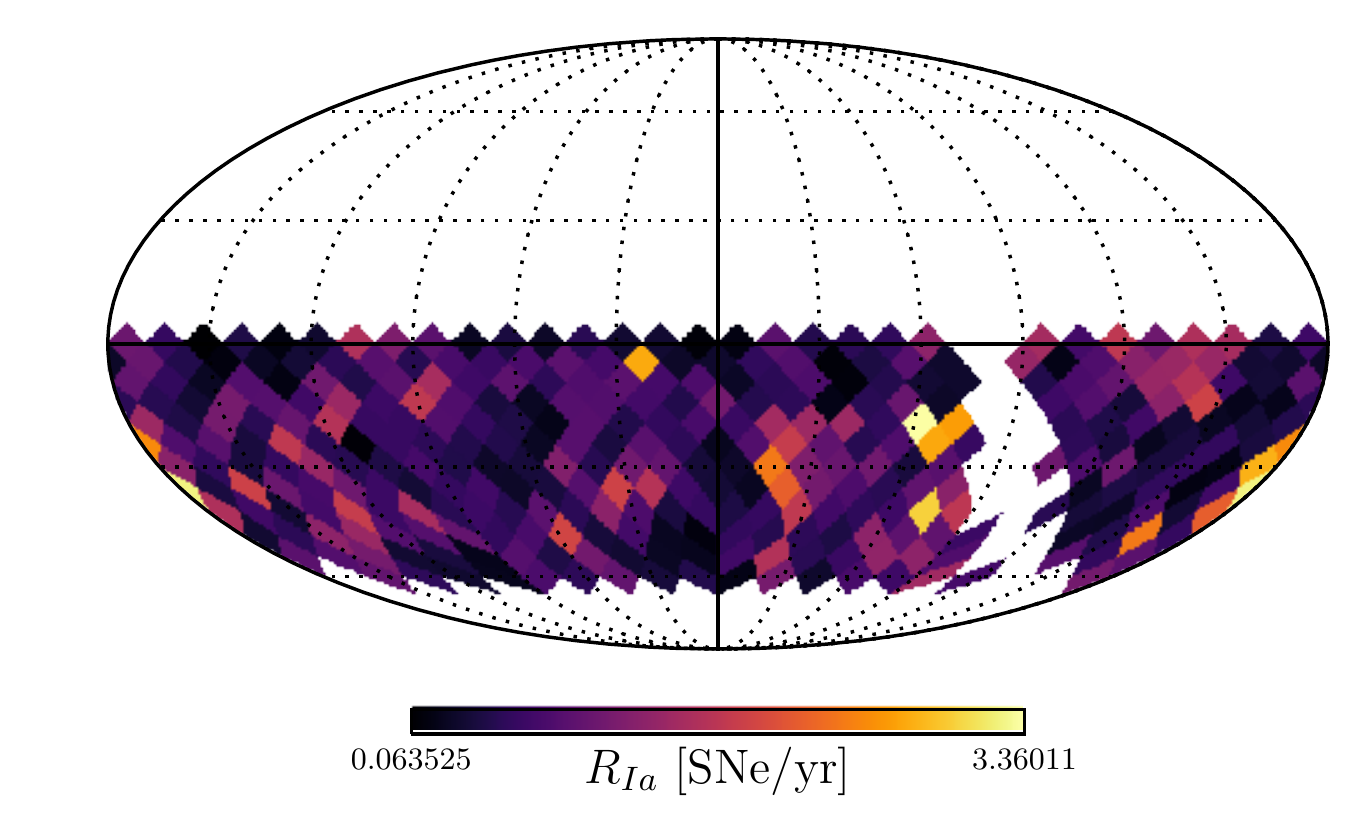}
	\caption{The type Ia supernova rate in each pixel in a shell at $\SI{140}{Mpc/h}-\SI{160}{Mpc/h}$, using the HEALPix pixelation scheme with $n_{side}=8$. Only the part of the sky that will be covered with the LSST main survey is shown -- the white stripe at the right corresponds to the galactic plane.}
	\label{fig:SNrate}
\end{figure}

The observed sources are distributed in redshift bins as illustrated in figure \ref{fig:distancebins}; these bins define a set of shells in which the velocity field will be measured. The velocity field in each shell is pixelized according to the HEALPix\footnote{\url{http://healpix.sourceforge.net}} pixelation scheme \cite{Gorski2005}, in which the number of pixels on the sphere is $N_\textrm{pix} = 12\cdot n_\textrm{side}^2$, and the parameter $n_\textrm{side}$ has to be chosen as an integer power of 2. An example of the pixelation can be seen in figure \ref{fig:SNrate}, which shows the average number of type Ia supernovae per year in each pixel in a shell at $140-\SI{160}{Mpc/h}$ for one of the observers. This also serves to show the geometry of the survey.

Since the volumes of the shells increase with distance for a given bin width, different resolutions can be obtained in different shells. We find that with a mean of approximately $5$ observations per pixel, there are relatively few empty pixels in the survey. To reach as high a pixel-resolution as possible, we divide the survey into three regimes with $n_\textrm{side}$ equal to 4, 8, and 16 respectively, as illustrated in figure \ref{fig:distancebins}, allowing us to resolve significantly higher multipoles in the outermost than in the innermost redshift shells.

Before we describe the procedure for obtaining the angular velocity power spectrum in each shell in section \ref{sec:AngularPowerSpectrum}, we specify a number of details concerning the mock observations in the paragraphs below. In order to determine the significance of different aspects of the analysis, and to facilitate comparison with other studies, we carry out a number of different analyses where the details are varied one by one. These are summarized in table \ref{tab:analyses}, where a naming scheme is also introduced. 

\begin{center}
	\begin{table}[h]
		\center 
		\begin{tabular}{ccccccc}
			\hline\hline
			Name & \specialcell{Observer\\positions} & \specialcell{Source\\selection} & \specialcell{Hosts with\\spectroscopy} & Binning & Geometry & \specialcell{Pixel\\weights} \\
			\hline 
			Ref & MW galaxies	& SN rates & All & $z$	& LSST & Uniform\\
			Full & MW galaxies & SN rates & All  & $z$	& Full sky & Uniform\\
			RndPos & Random & SN rates & All & $z$	& LSST & Uniform\\
			Hmw & MW galaxies & Halo masses & All & $z$ & LSST & Uniform\\
			Binr & MW galaxies & SN rates & All & $r$  & LSST & Uniform\\
			10k  & MW galaxies & SN rates & 10,000 & $z$ & LSST & Uniform\\
			Weighted & MW galaxies & SN rates & All & $z$ & LSST & \# Sources \\ 			
			\hline
		\end{tabular}
		\caption{This table specifies how several aspects of the measurements of the velocity field are varied in order to determine their individual significance. It also introduces a naming scheme for the different analyses.}
		\label{tab:analyses}
	\end{table}
\end{center}

\subsubsection*{The reference analysis}

In what we will refer to as the \textit{reference} analysis ('Ref'), we place 1000 mock observers in galaxies similar to the Milky Way, i.e. galaxies in which most of the stellar mass is found in the disk, and which are located in a dark matter halo with a mass in the range $10^{14}-\num{5e15}\,\di{M}_\odot$, which is similar to the mass of the Virgo Super Cluster. For each of these, we calculate the total type Ia supernova rate within a survey volume with the geometry of the LSST main survey, which is used to calculate the expected number of type Ia supernovae that will take place in a 10 years period. We then select this number of observed sources from the galaxy catalog, with the probability for selecting a given galaxy proportional to its present day type Ia supernova rate. This is equivalent to assuming that all type Ia supernovae within the survey volume are observed, and that a spectroscopic redshift is obtained for each of their host galaxies. At last, we bin the sources into shells according to their observed redshifts, as illustrated in figure \ref{fig:distancebins}, and calculate the angular power spectrum of the radial peculiar velocities in each shell, giving equal weights to all pixels containing at least one source.

Below, we describe how several aspects of the reference analysis are varied in a set of alternative analyses, each of which changes one aspect of the reference analysis, while keeping the rest unchanged.

\subsubsection*{Survey geometry}

In section \ref{sec:AngularPowerSpectrum}, we describe how the angular velocity power spectrum can be reconstructed from a survey covering only part of the sky. In order to check how this reconstruction affects the measured velocity field, we also carry out an analysis in which the full sky is observed ('Full'). 

\subsubsection*{Observer environment}

As mentioned above, the observers in the reference analysis are placed in locations similar to our own position in the universe. For the purpose of testing the effect of the environment on the measured velocity field, we include a different sets of observers, consisting of positions distributed randomly throughout the simulation volume ('RndPos'). Since voids take up a greater fraction of the simulation volume than overdense regions, random positions will tend to be located in underdense regions, implying that the set of random observers is genuinely different from the set of observers located in Milky Way-like galaxies.

\subsubsection*{Binning}

In the reference analysis, we distribute the observed sources in bins according to their measured redshifts, since in practice these are known much more precisely than the distances inferred from the type Ia supernova lightcurves. However, when aiming to measure the angular velocity power spectrum at different distances, this binning introduces an error, as some sources will be assigned to a different shell than the one corresponding to their true distance from the observer. The significance of this error is studied by including an analysis in which the observed objects are distributed in bins according to their true comoving distance rather than their measured redshifts ('Binr').

\subsubsection*{Weighting}

As the supernovae are not uniformly distributed over the sky, some pixels will contain a higher number of sources than others. In order to reduce noise stemming from the uncertainty in the measurements, pixels with a high number of observations should be given a larger weight in the determination of the velocity field. But this causes a bias, as the measured velocity field will then be dominated by regions of high density. We avoid this bias in the reference analysis by giving equal weights to all pixels, disregarding the number of observations they contain. We test how this affects the measurements by also including an analysis in which each pixel is given a weight proportional to the number of sources it contains ('Weighted').

\subsubsection*{Number of sources}

In the reference analysis, the number of selected sources is calculated from the total rate of type Ia supernovae within the observed region. As the average rate of type Ia supernovae in the main simulation is $\SI{1.1e-4}{SNe.yr^{-1}.h^{3}.Mpc^{-3}}$, this corresponds to approximately $\num{35000}$ supernovae within the survey geometry and in the distance range $40-\SI{256}{Mpc/h}$. However, the number of galaxies which can be used for a study of the velocity field such as the one described here does not only depend on occurrences of type Ia supernovae, but also on precise, i.e. spectroscopic, redshifts being available for the host galaxies. Since spectroscopy will not be carried out as part of the LSST survey, the potential for measuring the velocity field is dependent on spectroscopic redshifts obtained as part of other surveys. Luckily, redshifts are already known for many galaxies in the relevant redshift range, thanks to surveys such as SDSS \cite{SDSS2000} and, on the southern hemisphere, the 2dFGRS and 6dFGS surveys \cite{2dFGRS2001,6dFGS2009}. And since type Ia supernovae tend to occur in relatively bright galaxies, it seems likely that there will be a significant overlap between the galaxies in which a type Ia supernova is observed with the LSST and the galaxies already found in existing galaxy catalogs. In addition to this, a large fraction of the type Ia host galaxies will very likely be selected for spectroscopic follow-up after having been observed by the LSST.  

In view of these considerations, we believe it reasonable to base our measurements of the velocity field on the total number of type Ia supernovae identified in the mock surveys. However, to assess how much our results depend on these assumptions -- as well as on the total rate of type Ia supernovae -- we include an analysis where we limit the number of sources to 10,000 ('10k').

\subsubsection*{Distribution of sources}

By basing our mock observations on galaxy catalogs, and selecting sources based on their present day rate of type Ia supernovae, we aim to obtain a realistic distribution of sources for the subsequent measurements of the velocity field. To determine how much the scheme used to obtain the distribution of sources affect the measurements of the velocity field, we also perform an analysis based on the halo catalog at $z=0$, where we choose the observed halos with a probability proportional to their mass ('Hmw'). This also facilitates comparison with other studies, such as \cite{Hannestad2007}, which are based on dark matter halo catalogs. 


\section{The angular velocity power spectrum}
\label{sec:AngularPowerSpectrum}

In each shell, the radial peculiar velocities define a function on the sphere, $v_{r}({\unitvec{r}})$, where $\unitvec{r}$ is a unit vector identifying a point on the sphere. Such a function can be expanded in spherical harmonics, $Y_{lm}$, i.e.:
\begin{align}
v_{r}({\unitvec{r}}) & =\sum_{l=0}^{\infty}\sum_{m=-l}^la_{lm}Y_{lm}({\unitvec{r}}).
\label{eq:vr}
\end{align}
The power contained in fluctuations corresponding to multipole $l$ can be calculated from the coefficients $a_{lm}$ of the expansion. There are $2l+1$ modes for each $l$, and therefore the average power corresponding to a given value of $l$ is
\begin{align}
C_{l} & =\frac{1}{2l+1}\sum_{m}\fabs{a_{lm}}^{2}.
\label{eq:PS}
\end{align}
To obtain this power spectrum of the velocity field measured by each observer in each of the redshift bins shown in figure \ref{fig:distancebins}, we use the Healpy version of the HEALPix package, developed to carry out harmonic analyses of the CMB\footnote{See \url{http://healpix.sourceforge.net} for a discussion of the HEALPix conventions.}. The algorithm is based on the assumption that there is insignificant power in modes with $l>l_\textrm{max}$, as will be the case for a smoothed velocity field, so that the first sum in e.g. \ref{eq:vr} can be terminated at $l=l_\textrm{max}$: 
\begin{align*}
v_{r}({\unitvec{r}}) & =\sum_{l=0}^{l_\textrm{max}}\sum_{m=-l}^{l}a_{lm}Y_{lm}({\unitvec{r}}).
\label{eq:vrterm}
\end{align*}
The pixelation scheme is then used to define $N_\textrm{pix}$ locations in which the field is sampled, and the sampled values are used to estimate the coefficients of the expansion as
\begin{align*}
\hat{a}_{lm} & =\frac{4\pi}{N_\textrm{pix}}\sum_{p=0}^{N_\textrm{pix}-1}Y_{lm}^{*}(\unitvec{r}_p)v_r(\unitvec{r}_p),
\end{align*}
where $\unitvec{r}_p$ is the direction to the pixel denoted by $p$. From these estimated coefficients, the power spectrum can be obtained using equation \ref{eq:PS}. 

We do not expect that the analysis will be able to produce an accurate estimate of the dipole of the velocity field because of the limited sky cover. Therefore, as is recommended in the case of a cut-sky analysis in the HEALPix documentation, we remove both the best-fit monopole and dipole before calculating the rest of the power spectrum.

\subsection{Smoothing of the velocity field}

\begin{figure}[h]
	\centering
	\includegraphics[width=\textwidth]{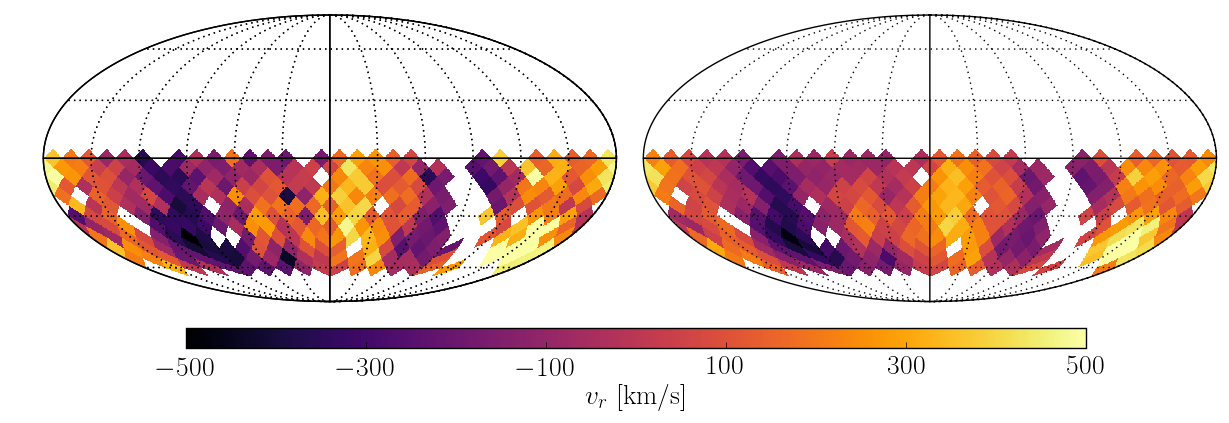}
	\includegraphics{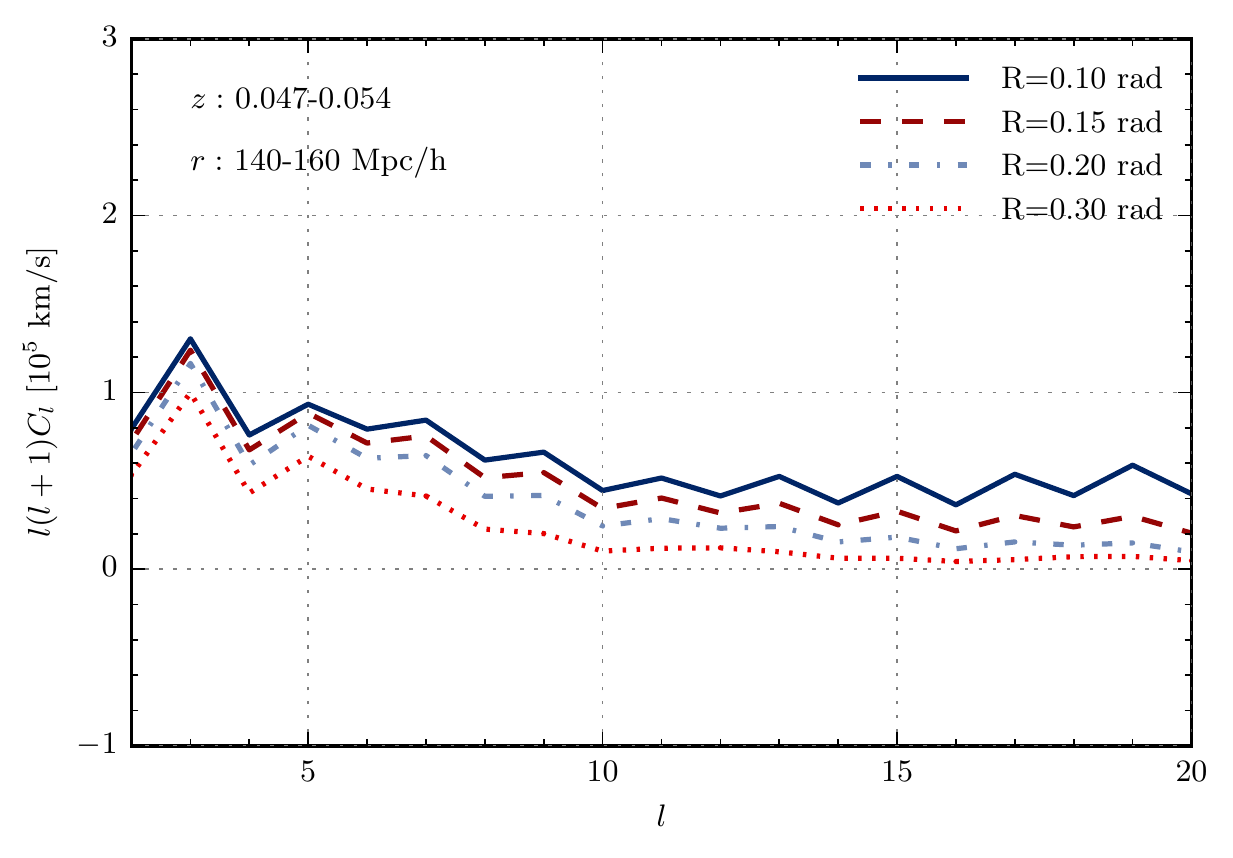}
	\caption{\textbf{Top:} Velocity field at $140-\SI{160}{Mpc/h}$ before (\textbf{left}) and after (\textbf{right}) smoothing with a Gaussian of FWHM $R = \SI{0.15}{rad}$.  \textbf{Bottom:} The mean power spectrum among the observers from a velocity field smoothed at the four different scales shown in the legend.}
	\label{fig:smoothing}
\end{figure}

Due to the survey geometry, the first multipoles of the expansion cannot be reliably estimated in the survey.  At the other end of the spectrum, the resolution, determined by the largest hole in the sky cover or the pixel size, sets an upper limit to the multipoles that can be determined. In order to reduce noise from sub-resolution variations, we use the Healpix smoothing-function to smooth the field with a Gaussian beam of full-width-half-maximum (FWHM) as specified in figure \ref{fig:distancebins}. To illustrate the effect of smoothing the field, we show in figure \ref{fig:smoothing} the mean power spectrum among the observers for one of the central bins ($140-\SI{160}{Mpc/h}$), after it has been smoothed with four different smoothing scales of respectively $\SI{0.1}{rad}$, $\SI{0.15}{rad}$, $\SI{0.2}{rad}$, and $\SI{0.3}{rad}$. The figure also shows the map of radial velocities for an example observer before and after smoothing.  

\subsection{Correcting for mode-coupling due to incomplete sky coverage}
\label{sec:MASTER}

\begin{figure}[h]
	\centering
	\includegraphics{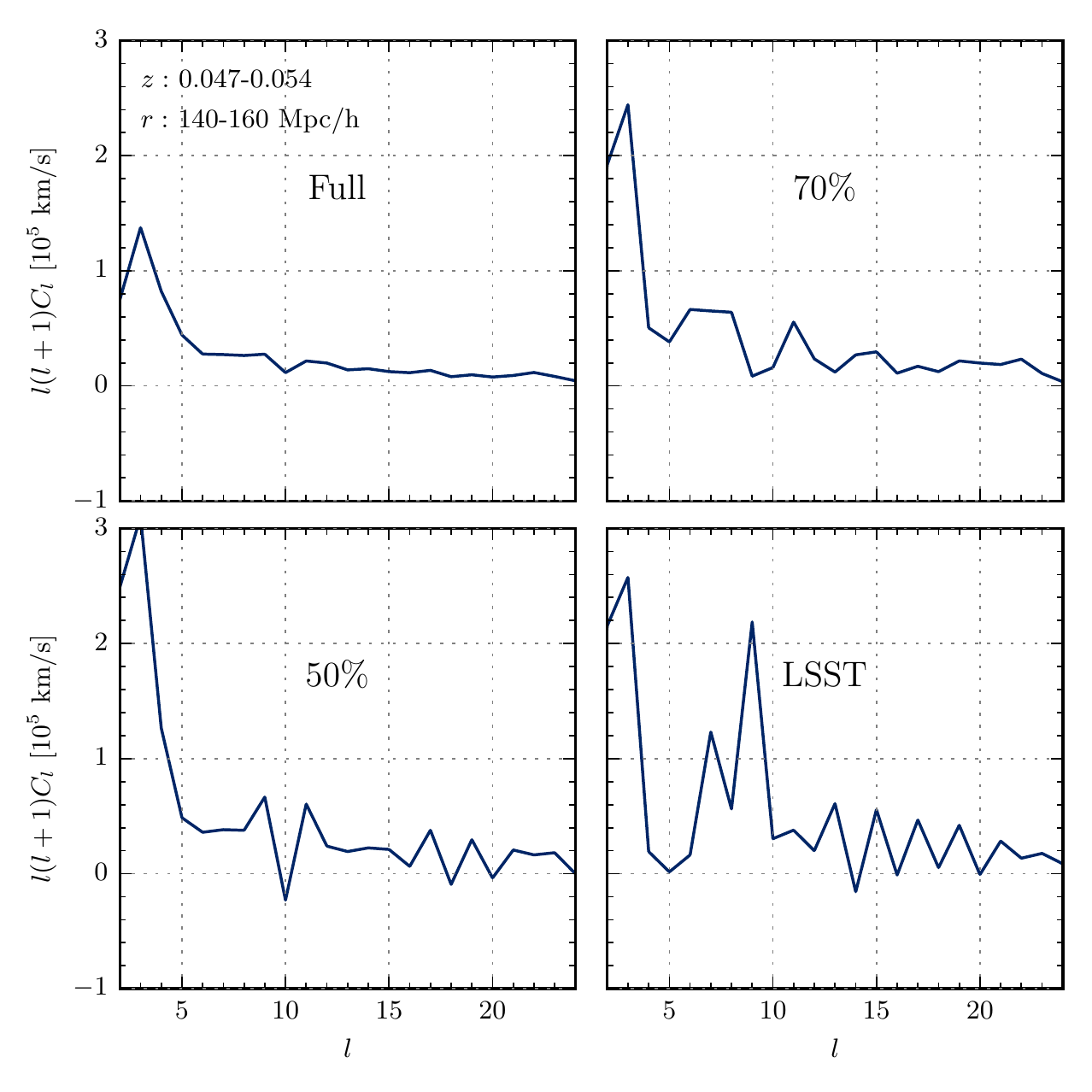}
	\caption{The angular power spectrum for a specific observer, based on four different survey geometries: observation of the full sky (\textbf{top left}), excluding a cap around the north pole so that only 70\% of the sky is observed (\textbf{top right}), excluding the entire northern hemisphere (\textbf{bottom left}), and using the survey geometry of the LSST survey (\textbf{bottom right}). }
	\label{fig:powerspectra_decreasing_skyfractions}
\end{figure}

We use the MASTER correction procedure, described in \cite{Hivon2002}, to correct for the missing sky coverage in the LSST survey geometry. Here, we describe the general idea, and sketch the mathematics behind the method.

Given a function defined on the entire sphere, such as the radial velocity field $v_r$, and a window function $W$ which describes where the function has been sampled (in this case the survey geometry of the LSST main survey), we can expand the part of the function that has been observed in spherical harmonics, $Y_{lm}$, as
\begin{align}
\tilde{a}_{lm} 
& = \int d\unitvec{r} v_r(\unitvec{r}) W(\unitvec{r}) Y^*_{lm}(\unitvec{r})\\
& = \sum_{l'm'}a_{l'm'}\int d\unitvec{r} Y_{l'm'}(\unitvec{r}) W(\unitvec{r}) Y^*_{lm}(\unitvec{r}).
\label{eq:pseudoalm}
\end{align} 
In the second line, the expansion of the full function in spherical harmonics as given in equation \ref{eq:vr} has been inserted. The integral in the second line describes how the different spherical harmonics couple to each other through the window function. By also expanding the window function in spherical harmonics, $W(\unitvec{r}) = \sum_{lm}w_{lm} Y_{lm}(\unitvec{r})$, this can be expressed as
\begin{align}
\int d\unitvec{r} Y_{l'm'}(\unitvec{r}) W(\unitvec{r}) Y^*_{lm}(\unitvec{r}) = 
\sum_{l_3m_3}w_{l_3m_3}\int d\unitvec{r} 
Y_{l_1m_1}(\unitvec{r})Y_{l_3m_3}(\unitvec{r})Y^*_{l_2m_2}(\unitvec{r}).
\label{eq:coupling}
\end{align} 
The integral in the last expression describes the coupling between a set of spherical harmonics, and this can be expressed in terms of Clebsch-Gordan coefficients.

The coefficients of the multipole expansion of the observed part of the function, as given in equation \ref{eq:pseudoalm}, define the so-called pseudo power spectrum:
\begin{align}
\tilde{C}_{l} \equiv \frac{1}{2l+1}\sum_{m=-l}^{l} |\tilde{a}_{lm}|^2. 
\label{eq:pseudoPS}
\end{align}
Ideally, we would like to be able to reproduce the power spectrum of the full sky, the $C_l$'s defined in equation \ref{eq:PS}, from the pseudo power spectrum and the coupling of modes introduced by the window function. However, this is not possible, since the available information tells us nothing about the unobserved part of the sky. But by assuming that the fluctuations of the field we are describing follows a Gaussian distribution, we can learn something about this underlying distribution. If the fluctuations of the field are pulled from a Gaussian, the $a_{lm}$'s must be so as well, and since the monopole is zero, $\langle a_{lm} \rangle = 0$, the variance and covariance of the coefficients can be calculated as $\langle a_{lm}a_{l'm'} \rangle = \delta_{ll'}\delta_{mm'}\langle C_{l}\rangle$. Here, $\langle C_{l}\rangle$ is the true variance of $a_{lm}$, or equivalently, the ensemble average of $C_l$ for all the fluctuation fields corresponding to a given cosmological model. By taking the ensemble average of equation \ref{eq:pseudoPS}, inserting the expressions from equations \ref{eq:pseudoalm} and \ref{eq:coupling}, and using the orthogonality relations of the Clebsh-Gordan coefficients and the diagonality of $\langle C_{l}\rangle$, one can show that the ensemble averages of the true and pseudo power spectra are related by
\begin{align}
\langle \tilde{C}_{l_1}\rangle = \sum_{l_2} M_{l_1l_2} \langle C_{l_2} \rangle. 
\end{align}
Here, $M_{l_1l_2}$ is the so-called MASTER matrix, given as  a sum over Clebsh-Gordan coefficients and the power spectrum for of window function, $\mathcal{W}_l$, as
\begin{align}
M_{l_1l_2} =\frac{2l_2+1}{4\pi} \sum_{l_3} (2l_3+1)\mathcal{W}_{l_3} 
  \begin{pmatrix}
  l_1 & l_2 & l_3\\
  0 & 0 & 0 &
  \end{pmatrix}^2,
\end{align}
where the matrix is the Wigner-symbol for the Clebsh-Gordan coefficients.  

In figure \ref{fig:powerspectra_decreasing_skyfractions}, we show how well the MASTER correction procedure works. The figure shows the measured angular velocity power spectrum in a shell at $140-\SI{160}{Mpc/h}$ for an example observer. The power spectrum has been determined using four different survey geometries: the first one covering the full sky (in which case no correction procedure is necessary), the second and third covering respectively 70\% and 50\% of the sky, and the last one corresponding to the geometry of the LSST main survey. For the three partial surveys, the MASTER method has been used to correct for the missing sky coverage. It is seen that the larger the unobserved part of the sky, the less reliable the multipole expansion becomes. However, as we will see in the next section, the mean of the recovered power spectra among all the observers is found to agree well with the spectra measured from the full sky, as the derivation of the MASTER method predicts. 

%
%
%

\section{Results and discussion}
\label{sec:Results}

\begin{figure}[h]
	\centering
	\includegraphics{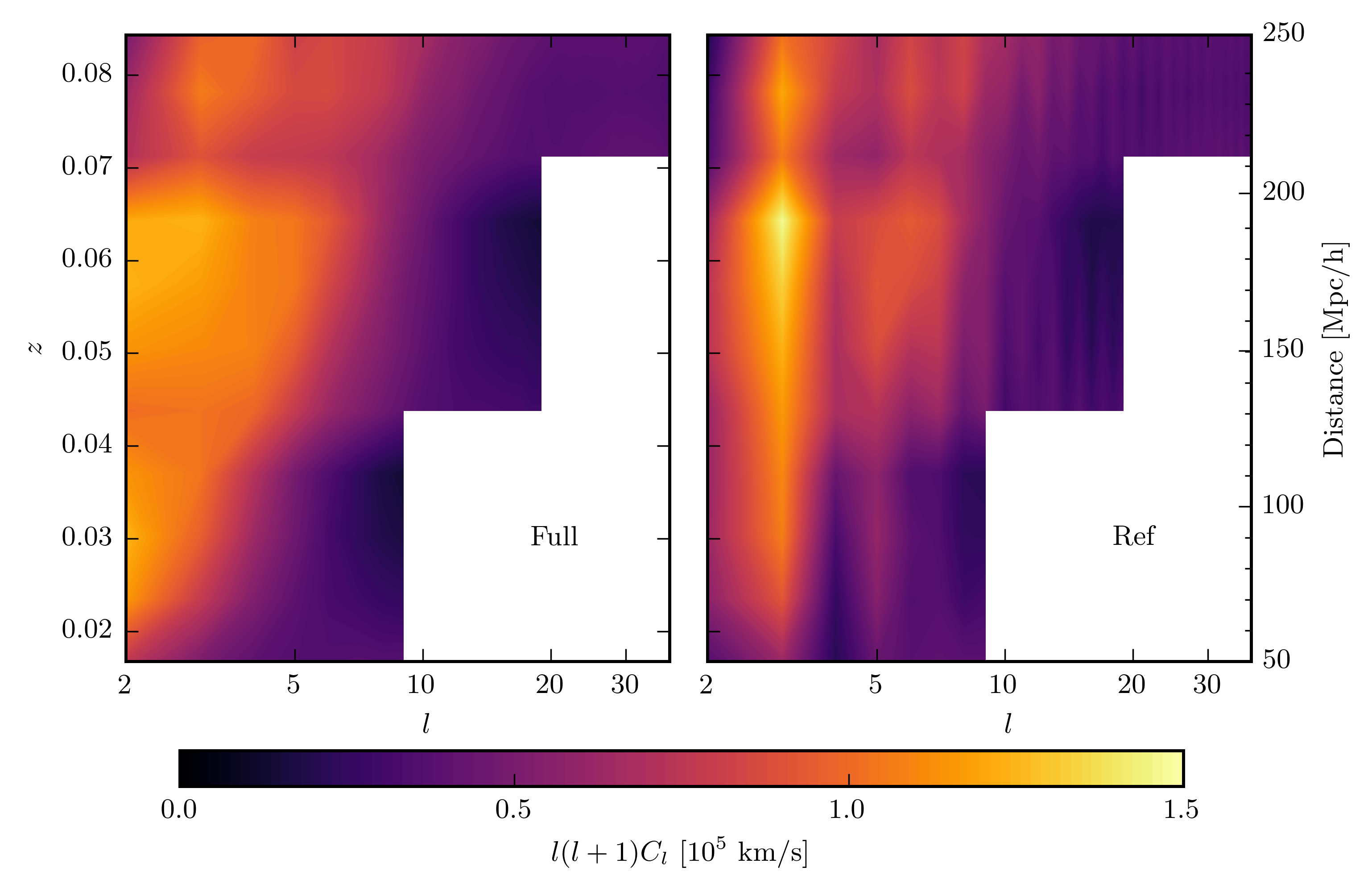}
	\caption{Contour plots showing the mean angular power as a function of both $z$ and $l$ in the case of full sky coverage (\textbf{left}) and in the reference scenario (\textbf{right}). Note that the $l$-axis is logarithmic.}
	\label{fig:3D}
\end{figure}

In the previous section, we described the MASTER correction procedure, and used it to recover the power spectrum from a single observer. In figure \ref{fig:3D}, the correction procedure is tested on the total set of 1000 reference observers, by comparing the mean of the measured velocity power spectra in the reference analysis with those measured by the same observers with full sky coverage. The mean of the power spectrum among the observers for all redshift bins and all $l$-values in the reference and full sky analyses are displayed as contour plots. Since a given length scale corresponds to increasing values of $l$ as the distance increases\footnote{The physical scale corresponding to a given value of $l$ can be estimated by remembering that each spherical harmonic oscillates $l$ times over the sphere, so the wavelength of an oscillation in a shell at distance $r$ can be obtained as $\lambda=2\pi r/l$.}, we expect the peak of the power spectrum to move towards higher $l$ for increasing redshift, which is indeed the most prominent feature seen in the mean power spectrum obtained from the full sky. The same pattern is seen in the power spectrum reconstructed from surveys with the LSST geometry, revealing a good overall agreement between the full and the reconstructed power spectra, although some spurious oscillations are introduced by the reconstruction.

\begin{figure}[h]
	\centering
	\includegraphics{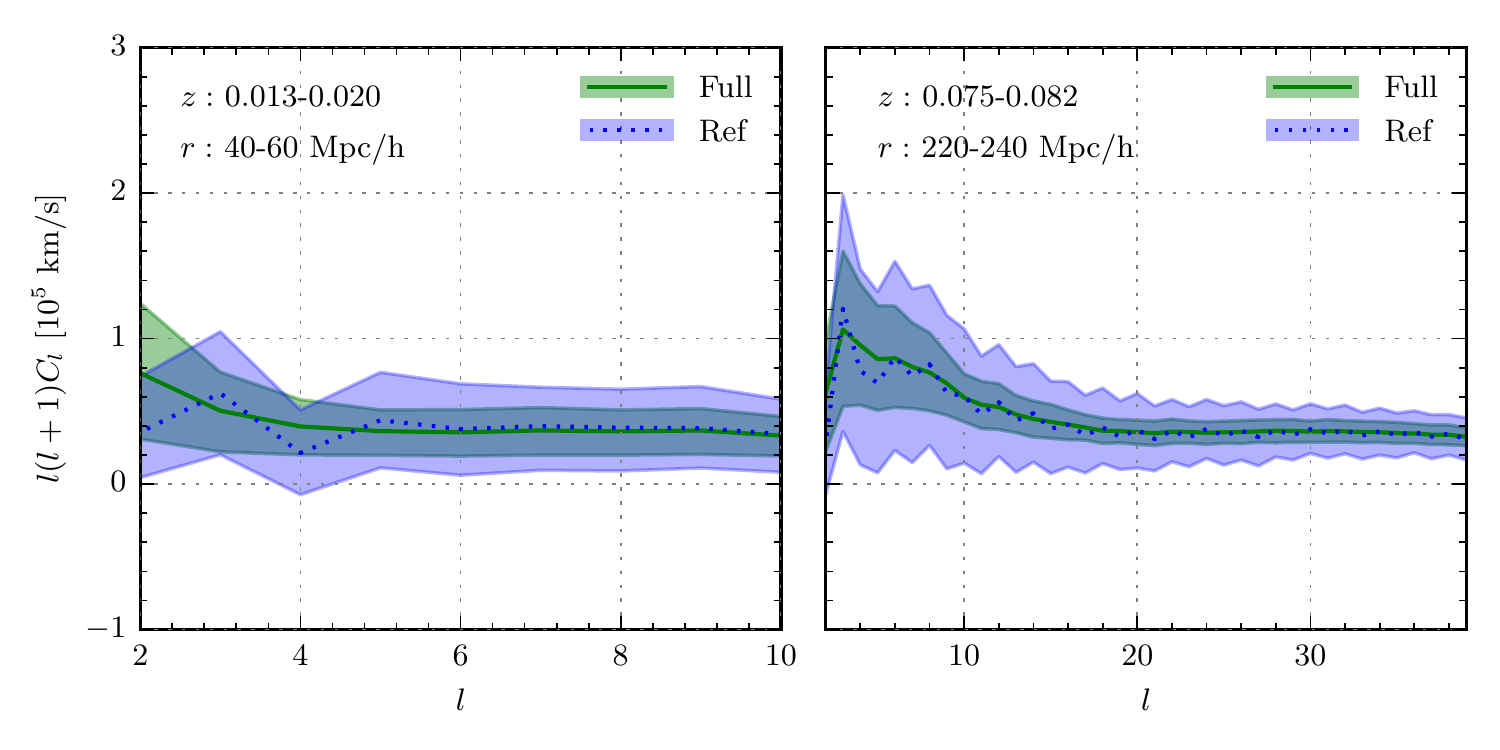}
	\caption{Comparison between the angular velocity power spectra in the reference and full sky analyses. The figures show the mean and spread in the spectra in two different bins, containing sources with redshifts from $z=0.013$ to $z=0.02$ and from $z=0.075$ to $0.082$, respectively. Assuming the redshifts to be purely cosmological, this corresponds to distances of $40-\SI{60}{Mpc/h}$ and $220-\SI{240}{Mpc/h}$. The mean of the power spectra among the observers for the full sky analysis is shown with a full, green line, and the 68\% confidence interval is shown in green shadings. The mean and spread in the reference analysis are shown with a blue dotted line and blue shadings for comparison.}
	\label{fig:ref_vs_full}
\end{figure}

\begin{figure}[h]
	\centering
	\includegraphics{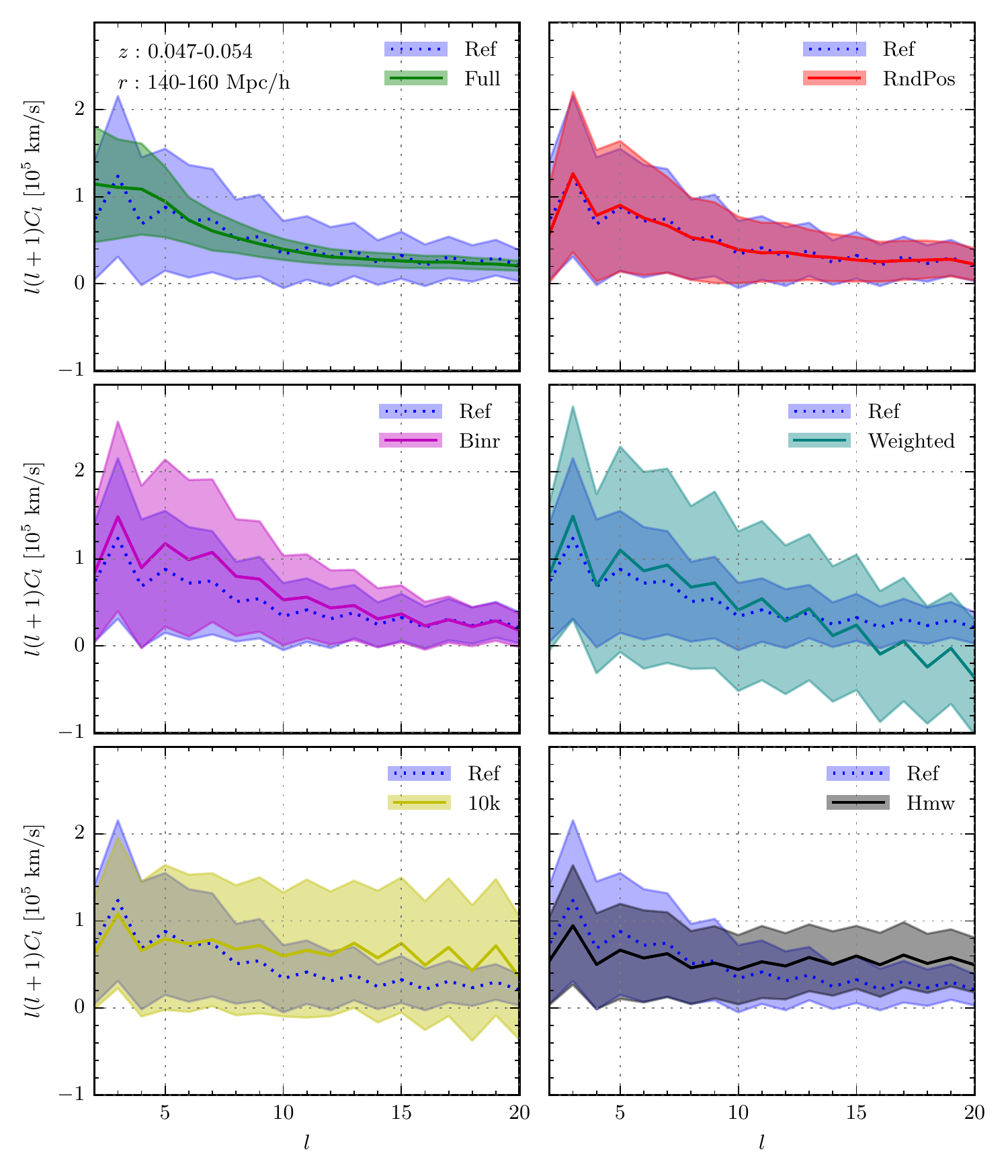}
	\caption{In this figure, the mean and spread in the angular velocity power spectra for sources with redshifts between $z=0.047$ and $z=0.054$, corresponding to distances of $140-\SI{160}{Mpc/h}$ if the redshifts are assumed to be cosmological, are shown for each of the analyses listed in table \ref{tab:analyses}. In each case, the mean of the power spectrum among the observers is shown with a full line, and the 68\% confidence interval is shown in colored shadings. The mean and spread in the reference analysis are shown with a blue dotted line and blue shadings for comparison.}
	\label{fig:cmp}
\end{figure}

These conclusions can also be drawn by studying the comparison between the reference and full sky analyses in figure \ref{fig:ref_vs_full} and the top left panel of figure \ref{fig:cmp}. In these plots, the mean of the power spectra as well as the 68\% confidence interval in the two cases are shown for three different redshift shells. Again, good agreement is found between the mean of the angular power spectra among the observers in the two cases, as predicted in the derivation sketched in section \ref{sec:MASTER}. There is a large increase in the spread in the recovered power spectra, as compared to that measured from the full sky, which is expected, since the recovered power spectra are constructed from a much smaller amount of information.



In figure \ref{fig:cmp}, the reference analysis is also compared to each of the other analyses mentioned in section \ref{sec:ObserversAndObservations} and summarized in table \ref{tab:analyses}, illustrating how different aspects of the analysis affect measurements of the power spectrum. 

From the comparison between the power spectra determined by Milky Way-like observers and observers placed at random positions throughout the simulation volume, we find that the observer environment makes virtually no difference. In contrast, the binning of the observed sources has a significant effect, where binning in distance rather than redshift is seen to result in an increase of power at low and intermediate multipoles. This can be understood from the fact that there is a higher probability for sources with a large peculiar velocity to get assigned to a shell not corresponding to the actual physical distance to the source. Note, however, that the difference vanishes at high multipoles. This implies that the power at high multipoles stem from smaller radial velocities; we have confirmed this by artificially removing galaxies with a large radial velocity and noting that this has a large effect on the power spectrum at small and intermediate multipoles, and negligible effect at large multipoles.

The remaining three analyses differ from the reference analysis in how the observations are distributed spatially and what weights different regions receive. When greater weights are given to pixels containing many sources, the correction procedure still does a descent job of recovering the mean of the power spectrum at small and intermediate multipoles, although the spread in the measured power spectra increases considerably. When only 10,000 sources are used, the maximal $l$-value for which the power spectrum can be determined decreases significantly, as expected. A similar trend is observed at large multipoles in the analysis based on the halo catalog. This implies that small scales are not sufficiently resolved in this analysis, which would be the case if the sources were localized in fewer areas, resulting in a less uniform distribution. We have confirmed this upon closer inspection of the source distribution in each of the two cases. Lastly we note that the sources pulled from the halo catalog results in somewhat less power at small multipoles.

In this study, we have not considered the error stemming from the uncertainty in the measurements of the distances and redshifts of galaxies, only the uncertainty from the distribution of supernovae and the survey geometry, as well as cosmic variance. In \cite{Hannestad2007}, the uncertainty stemming from the distribution and binning of sources is referred to as the "geometric error", and according to their figure 2, these dominate for multipoles with $l \lesssim 10$. By comparing our results for the uncertainty in the measured power spectrum to figure 3 in 
\cite{Hannestad2007} for $l \lesssim 10$, we conclude that our results are in agreement with their finding of the precision with which the angular velocity power spectrum can be determined, and therefore, their conclusions are backed by this study. This implies that their result from the likelihood analysis applies for the distribution of supernovae and survey geometry considered in this study as well. As they find that the angular velocity power spectrum can be used to determine $\sigma_8$ with 3-5\% accuracy at 95\% confidence, we expect the velocity field from type Ia supernovae to become an independent probe of this and other parameters associated with the matter distribution in the near future. We will explore this further in a future paper.

\FloatBarrier
\section{Conclusions}
\label{sec:Conclusions}

We have studied the potential for measuring the power spectrum of radial peculiar velocities from type Ia supernovae in an LSST-like sky survey. The analysis shows that we can reasonably correct for the effects of missing sky cover due to the survey geometry, though with a penalty in the form of a large increase in the uncertainty of the measured spectrum. By carrying out the analysis for observers in different environments, we have found that the observer position does not cause any significant bias in the measured velocity power spectrum. On the other hand, we have seen that the procedure used for selecting the supernovae from which the velocity field is measured has a relatively large effect, implying that analyses based on inaccurate distributions of supernovae will produce somewhat biased results. Likewise, the choice for how the measurements are binned can have a significant effect on the results. 

Due to the close relationship between the matter distribution and the peculiar velocity field, the method studied here can be used to measure a set of fundamental cosmological parameters. By comparing our results with those obtained in \cite{Hannestad2007}, we conclude that the velocity field from type Ia supernovae will be able to provide estimates of $\sigma_8$ with a precision comparable to what can be obtained from measurements of the galaxy power spectrum or weak lensing (see for example \cite{Giannantonio2012,Huff2013}). As the most recent value determined from the CMB points to a much higher value of this parameter than analysis of weak lensing alignment of galaxies \cite{Planck2015,Heymans2013}, this might become valuable as a complimentary way to measure this parameter in the near future.


\section{Acknowledgements}

Computer resources from the Centre for Scientific Computing Aarhus have been used for the analyses presented in this paper. A large part of the analysis has been carried out using the HEALPix \cite{Gorski2005} package. We thank Andrew Benson, who has been helpful in solving problems with the installation and use of Galacticus.

\bibliographystyle{utcaps}
\bibliography{LSSTVelocityField}

\end{document}